\documentclass[12pt]{article}
\usepackage{amsmath,amssymb}
\usepackage{graphicx,psfrag,epsf}
\usepackage{enumerate}
\usepackage{natbib}
\usepackage{caption}
\usepackage{subcaption}
\usepackage[colorlinks, citecolor = blue, linkcolor = blue]{hyperref}
\usepackage{centernot}
\usepackage{cite}
\usepackage{authblk}
\usepackage{url} 
\usepackage{bm}
\usepackage{floatrow}

\newcommand{\blind}{1}
\newcommand{\iid}{\stackrel{iid}{\sim}}
\newcommand{\E}{\mathbb{E}}
\newfloatcommand{capbtabbox}{table}[][\FBwidth]
\newcommand{\indep}{\raisebox{0.05em}{\rotatebox[origin=c]{90}{$\models$}}}

\usepackage{tikz}
\usetikzlibrary{shapes,arrows}

\begin{document}
\tikzstyle{block} = [rectangle, draw, fill=blue!20, 
    text width=5em, text centered, rounded corners, minimum height=4em]
\tikzstyle{line} = [draw, -latex']
\tikzstyle{cloud} = [draw, ellipse,fill=red!20, node distance=3cm,
    minimum height=2em]

\def\spacingset#1{\renewcommand{\baselinestretch}%
{#1}\small\normalsize} \spacingset{1}


\if1\blind
{
  \title{\bf Statistical Inference on Tree Swallow Migrations with Random Forests}
 \author[1]{Tim Coleman}
\author[1]{Lucas Mentch}
\author[2]{Daniel Fink}
\author[2]{Frank A. La Sorte}
\author[3]{David W. Winkler}
\author[4]{Giles Hooker}
\author[2]{Wesley M. Hochachka}
\affil[1]{\small University of Pittsburgh, Department of Statistics}
\affil[2]{\small Cornell University, Lab of Ornithology}
\affil[3]{\small Cornell University, Cornell University, Department of Ecology and Evolutionary Biology}
\affil[4]{\small Cornell University, Department of Statistics}
  \maketitle
} \fi

\if0\blind
{
  \bigskip
  \bigskip
  \bigskip
  \begin{center}
    {\LARGE\bf Statistical Inference on Tree Swallow Occurrence with Random Forests}
\end{center}
  \medskip
} \fi


\begin{abstract}
\noindent Bird species' migratory patterns have typically been studied through individual observations and historical records.  In recent years however, the eBird citizen science project, which solicits observations from thousands of bird watchers around the world, has opened the door for a data-driven approach to understanding the large-scale geographical movements.  Here, we focus on the North American Tree Swallow (\textit{Tachycineta bicolor}) occurrence patterns throughout the eastern United States.  Migratory departure dates for this species are widely believed by both ornithologists and casual observers to vary substantially across years, but the reasons for this are largely unknown.  In this work, we present evidence that maximum daily temperature is predictive of Tree Swallow occurrence.  Because it is generally understood that species occurrence is a function of many complex, high-order interactions between ecological covariates, we utilize the flexible modeling approach offered by random forests.  Making use of recent asymptotic results, we provide formal hypothesis tests for predictive significance of various covariates and also develop and implement a permutation-based approach for formally assessing interannual variations by treating the prediction surfaces generated by random forests as functional data.  Each of these tests suggest that maximum daily temperature is important in predicting migration patterns. 
\end{abstract}
\noindent%

\newpage
\spacingset{1} 

\section{Introduction}
\label{sec:intro}
Tree Swallows (\textit{Tachycineta bicolor}) are migratory aerial insectivores. In a recent breeding season study, \citet{winkler2013temperature} suggested that maximum daily temperature during the breeding season had a significant effect on the abundance of the flying insects that are the primary food source of Tree Swallows. This local-scale study conducted in upstate New York established how cold snaps,  defined as two or more consecutive days when the maximum temperatures did not exceed 18.5$^\circ$C, can result in a diminished food supply, thereby suggesting an indirect link between lower temperatures and lower fledgling success. Other work supports the hypothesis that migratory birds, like Tree Swallows, have breeding patterns that are affected by climate change, \citep{dunn1999climate,hussell2003climate}. While these papers focus heavily on breeding success and food availability, in this work, we investigate the associations of temperature on regional and local patterns of species occurrence during the autumn migration.

Most ornithological studies rely on controlled, local or regional level studies during a single season of the year, limiting the spatial and temporal scope of the analysis. The eBird project \citep{sullivan2009ebird, sullivan2014ebird} hosted by the Cornell University Lab of Ornithology is a global bird monitoring project that allows for analysis on a much larger scale. This citizen science project compiles crowd-sourced observations of bird sightings, opening the door for a more data-driven approach to formally investigate scientific questions of interest. The eBird project harnesses the efforts of the bird-watching community by encouraging bird-watchers (birders) to record checklists of the species they encountered on each outing. These data have been used in a range of applications such as describing bird distribution across broad spatiotemporal extents \citep{fink2010spatiotemporal, Fink2018}, prioritizing priority habitat to conservation \citep{Johnston2015}, and identifying continental-scale constraints on migratory routes \citep{LaSorte2016}.

We study Tree Swallow populations during the autumn migration. During this time, the species are believed to be \textit{facultative migrants}. Facultative migration is an opportunistic migration strategy where individuals migrate in response to local conditions, such as the prevailing food supplies or weather conditions. Specifically, we study the low elevation New England / Mid-Atlantic Coast stretching north from the Chesapeake Bay to Boston, known as Bird Conservation Region 30 (BCR30) \citep{Sauer2003} that forms the northern extent of the Tree Swallow winter range in Eastern North America. 

Anecdotal accounts from bird watchers in this region suggest that Tree Swallows inhabit this region for prolonged autumn periods only during relatively warm winters.  Though never formally documented or proven, in the years 2008 and 2009 it was widely believed in the ornithological community that the species did not linger in the region as late into the autumn as usual. Alternatively, mortality in the northern parts of the range in those years may have been higher. Thus, our primary objectives in this work are twofold:  (1)  to formally test whether the temporal pattern of Tree Swallow occurrence during the autumn migrations of 2008 and 2009 were substantially different than what would be expected during a typical autumn migration and (2) if there are differences, to investigate the association between local-scale patterns of occurrence and daily maximum temperature across broad geographic extents. 

\subsection{Challenges in Modelling Tree Swallows}
While the ecological questions in the previous section are relatively straightforward to pose, providing accurate answers and provably valid statistical inference is challenging. In general, we expect that as daily temperatures decrease, the occurrence rate of Tree Swallows should also decrease as the species gravitates towards regions with more plentiful food or suffers higher mortality where food availability has been driven down by cold maximum temperatures. However, there are many strong sources of variation affecting the observed local-scale spatiotemporal patterns of species occurrence during the migration that can modify and mask the local-scale predictive utility of temperature. Ecological patterns of local-scale occurrence are affected by elevation, land cover types (e.g.\ open fields vs forests), and weather. Because of the difficulty finding and identifying birds in the field, variation in detection rates further complicates modeling and inference about the underlying ecological processes. Based on previous work (see, for example, \citet{Zuckerberg2016}), we expect such associations to appear as complex, high-order interactions among the available covariates and that many of these associations and interactions will vary throughout the autumn migration. 

Thus, one of the main analytical challenges is to develop models that can exploit rich covariate information to account for varied and complex sources of variation while facilitating statistical inference about potentially complex, local-scale associations. The large amount of available data, together with the presence of both nonlinear and high-order interactions, complicates the use of most traditional parametric and semiparametric models for this task. Thus, we rely on the more flexible alternative offered by random forests \citep{breiman2001random}.  Random forests have a well-documented history of empirical success and are considered to be among the best ``off-the-shelf'' supervised learning methods available \citep{fernandez2014we}. This strong track record of predictive accuracy makes them an ideal ``black-box" model for complex natural processes. Furthermore, tree-based methods have also proven very successful in other eBird projects \citep{Robinson2017,Fink2018}. 

Though black-box model are not easily amenable to statistical inference, recent asymptotic results from \citet{mentch2016quantifying, Peng2019} and \citet{Wager2017} on the distribution and variance estimation of predictions resulting from RF models provide a formal statistical framework for addressing our primary questions of interest. Moreover, as we demonstrate, traditional non-parametric inferential procedures can also be used to help draw inferences from these complex models. 

In this paper, we begin with a brief overview of the data and available covariate information in Section \ref{sec:data}.  In Section \ref{sec:prelim_mod}, we provide further evidence for use of random forests to answer the questions posed earlier. We then construct preliminary RF models to assess the influence of temperature and produce maps of prediction differences between models to understand the spatial patterns in the association with maximum daily temperature. In Section \ref{sec:glob}, we develop a permutation-style test to investigate how unusual the 2008 and 2009 migration patterns appear to be by treating the RF predictions over time as functional data.  Finally, In Section \ref{sec:loctest}, we make use of recent asymptotic results to test the significance of maximum daily temperature at a variety of local test locations throughout the region of interest, BCR30. Throughout this work, the predictive associations uncovered should not be interpreted as causal effects.  Indeed, due to the structure of the tree swallow data, it is more likely that the causal relationship between maximum daily temperature and \texttt{occurrence} is indirect, as maximum daily temperature affects food availability or other local resources upon which Tree Swallows depend.. However, we recognize the important work done in extending random forests to estimation of causal effects, and as such, a section implementing the causal forests of \citet{Wager2017, Athey2019} is provided in the supplementary material.

\section{Data Overview}
\label{sec:data}
The eBird data is accumulated on a per-birder outing basis. During each outing, the birder records the species of birds observed. Each species observed is recorded as a presence observation while unobserved species are marked absent. The outing is then referenced with environmental, spatial, temporal, and user information.  This last set of predictors is included in order to account for variation in detection rates, a potential confounder when making inference about species distributions. Our outcome of interest is the probability that at least one Tree Swallow is observed given the spatial, temporal, and detection process information. We refer to this probability as \textit{occurrence}, that is
\[
O= P(\text{Tree Swallow is Observed} \ |\ X = x) 
\]
where $X$ denotes the covariate information. Because we are interested in the eastern autumn migration, we restrict our attention to eBird observations located in the BCR30 region that were recorded on or after the 200th day of the year between the years 2008-2013. In total, the full dataset contains 173002 observations on 30 variables, with occurrences of tree swallows in 10.8\% of the observations.

 
Spatial information is captured by land cover and elevation data. To account for habitat-selectivity each eBird location has been linked to the remotely-sensed MODIS global land cover product (MCD12Q1) \citep{friedl2010modis}.  Here we use the 2011 MODIS land cover data as a static snapshot of the landcover. These landcover predictors were associated with eBird observations collected from 2004 to 2012. Finally, we use the University of Maryland (UMD) classification scheme \citep{hansen2000global} to classify each 500m $\times$ 500m pixel (25 hectare) as one of 14 classes, including classes such as water, evergreen needleleaf forest, and grasslands. We summarized the land cover data as the proportion of each land cover class within a 3.0km $\times$ 3.0km (900 hectare) pixel centered at each location using FRAGSTATS \citep{mcgarigal2012fragstats}. 

Temporal information is included at three resolutions. At the finest temporal resolution, the time of the day at which the observation was made is used to model variation in availability for detection; e.g., diurnal variation in behavior \citep{diefenbach2007incorporating} may make species more or less conspicuous. For our purposes, we restrict our attention to the day of year (\texttt{DoY}) and the year itself, corresponding to our interest in anomalies in the fall migration.

Temperature data was collected from the DayMet project, hosted by Oak Ridge National Lab \citep{ornl2013temp}. The data includes daily maximum (\texttt{max\_temp}), minimum, and mean temperature for each day in the training period. We also estimated an expected daily maximum temperature for each day by taking the mean daily maximum temperature for each eBird location from 1980-2007. The anomaly relative to this expected maximum (\texttt{max\_temp\_anomaly}, defined as \texttt{max\_temp} minus the 1980-2007 normal \texttt{max\_temp}) is of particular interest since \texttt{max\_temp} alone is strongly correlated with \texttt{DoY}. Each eBird location is further associated with the 30m gridded elevation from the ASTER Global Digital Elevation Model Version 2. 

Finally, there are three user effort variables included in the model to account for variation in detection rates: the hours spent searching for species (\texttt{eff\_hours}), the length of transects traveled during the search (\texttt{eff\_dist}), and the number of people in the search party (\texttt{n\_obs}).  In addition, an indicator of observations made under the ``traveling count" protocol was included to allow the model to capture systematic differences in species detection between the the counts recorded by traveling and stationary birders.

\section{Preliminary Models}
\label{sec:prelim_mod} 
Random forests are often exceptionally accurate and robust supervised learners \citep{fernandez2014we}.  Our primary goal is inference about associations between environmental variables and tree swallow migrations.  However, to trust such inference, we must trust that the model selected is able to accurately capture the complex underlying mechanisms. We thus begin our study by determining whether there are notable gains in accuracy by using random forests relative to alternative, more straightforward statistical models.  To answer this using the eBird data, we use cross validation (CV) to measure the predictive accuracy of a variety of popular modeling techniques for binary outcomes. Details and specifics of this tuning are provided in the supplementary materials. Even without tuning, the off-the-shelf random forest model attains the lowest CV errors.

\subsection{Inspecting Annual Migration Differences}
\label{subsec:RFheuristic}

Recall that the motivation for this study is a widely perceived but yet unproven difference in Tree Swallow autumn distribution patterns in the years 2008 and 2009.  In particular, it was believed that Tree Swallows had remained in the northern regions for a shorter period in the fall during 2008-2009, but the ornithological community was unsure of the mechanism(s) behind this earlier departure/decline. We begin by partitioning our (training) data $\mathcal{D}$ into $\mathcal{D}_{08-09}$ and $\mathcal{D}_{10-13}$, corresponding to each year grouping.  From \texttt{DoY} 200 onward, 100 points were removed at random from $\mathcal{D}$ to serve as a validation set. 

\begin{figure}
\begin{center}
     \includegraphics[width = 0.6\textwidth]{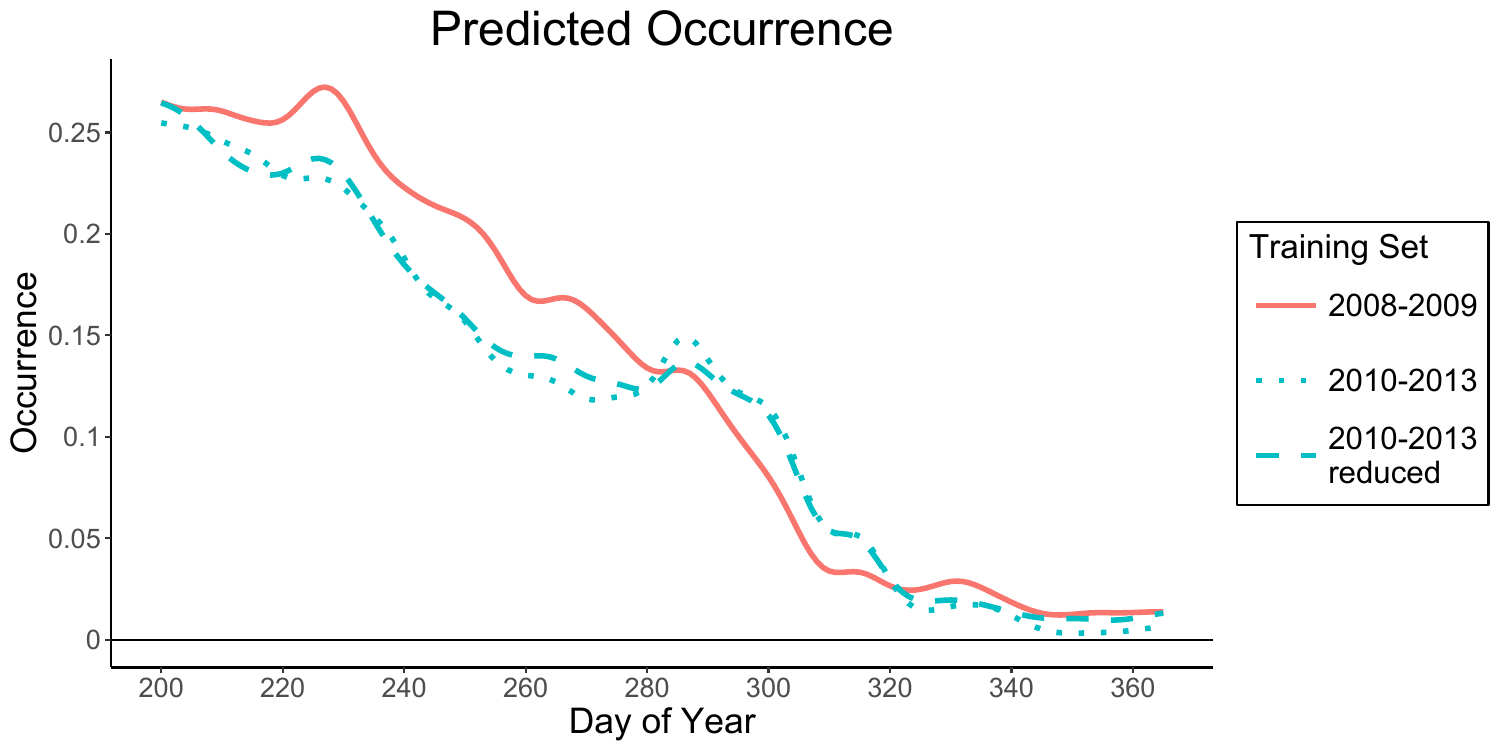}
\caption{Kernel smoothed predicted occurrence by random forests trained on $\mathcal{D}_{08-09}$, $\mathcal{D}_{10-13}$, and a subsample of $\mathcal{D}_{10-13}$. }
\label{RF_TSO_PE}
\end{center}
\end{figure}

We first construct a RF on each of these temporally divided training sets. It is important to note, however, that the eBird project has grown substantially in popularity since its inception in 2002 and thus later years contain many more observations than earlier years. In particular, $\mathcal{D}_{08-09}$ contains a total of 21,907 observations, while $\mathcal{D}_{10-13}$ contains 151,095.  Because a RF trained on a larger dataset may be more stable, any differences observed between predictions generated by the two datasets may be partially explained by the difference in data sizes.  To account for this, we also selected (uniformly at random, without replacement) a subsample of size 21,907 from the $\mathcal{D}_{10-13}$ training data and with it, constructed a third RF.  Predictions were made at all points in the validation set, and averaged by day.  The results are shown in Figure \ref{RF_TSO_PE}. 

From Figure \ref{RF_TSO_PE}, we see that the RF trained on $\mathcal{D}_{08-09}$ predicts a larger occurrence until approximately \texttt{DoY} 285, after which the 2010-2013 predictions are higher until approximately \texttt{DoY} 320, from which point the differences appear negligible.  This seems to support the hypothesis that in 2008-2009, Tree Swallows remained in northern regions longer before departing more quickly.  Importantly, the predictions from the RF trained on the reduced data from 2010-2013 forest differs only very slightly from those generated by the RF trained on the full $\mathcal{D}_{10-13}$ data, suggesting that the more substantial departures observed between predictions generated by RFs trained on  $\mathcal{D}_{08-09}$ and $\mathcal{D}_{10-13}$ have little to do with the differing training sample sizes.

As noted earlier, the temporal structure of the data complicates any kind of causal analysis. Based on \citet{winkler2013temperature}, we know that daily maximum temperature can act as a proxy for food availability, via insect abundance, a suspected causal ancestor of occurrence. As such, there is likely not a (direct) causal relationship between daily maximum temperature and occurrence.


\section{Testing for Regional Differences in Occurrence}
\label{sec:glob}
We now devise and implement a permutation test to explicitly assess whether the prediction curves in Figure \ref{RF_TSO_PE} exhibit differences that could plausibly be due to chance.  Here we consider \emph{regional} hypotheses, meaning that we investigate differences in species occurrence throughout the entire BCR30 region as opposed to at a specific location or set of locations within that region.

\subsection{Testing Procedure and Data}
\label{subsec:glob}

Our strategy here is to use a permutation test to investigate hypotheses about the distribution of occurrence in the 2008-2009 and 2010-2013 groupings. Permutation tests, in addition to maintaining exact control of the Type I error rate for distributional hypotheses, have the advantage of being completely distribution free, regardless of the test statistic used. If we let $D_{08-09}(\bm{X}, Y)$ denote the joint distribution of the covariates and occurrence for the 2008-2009 data, and similarly define $D_{10-13}(\bm{X}, Y)$, we then want to test
\begin{equation} \label{H0_dist}
\begin{split}
\text{H}_0:& \ D_{08-09}(\bm{X}, Y) = D_{10-13}(\bm{X}, Y)  \\
\text{H}_1:& \ D_{08-09}(\bm{X}, Y) \neq D_{10-13}(\bm{X}, Y) .
\end{split}
\end{equation}
To account for differing training set sizes and also in the interest of both computational efficiency and being conservative in our testing procedures, we now construct a reduced training set from the $\mathcal{D}_{10-13}$ data containing the same number of observations as $\mathcal{D}_{08-09}$.  We construct this reduced set by taking each observation in $\mathcal{D}_{08-09}$ and drawing a radius around it in both space (0.2 decimal degrees in both latitude and longitude, an area of approximately 352 km$^2$) and time (2 days).  We then locate all observations from $\mathcal{D}_{10-13}$ within this radius and select from these an observation uniformly at random, without replacement. This produces a ``nearest neighbor'' training set, $\mathcal{D}^{NN}_{10-13}$,  with roughly the same spatiotemporal distribution of observations, allowing us to more closely examine the influence of the other covariates on the functional observations. By enforcing spatio-temporal uniformity between the datasets, we are controlling for differences in eBird user behavior between the two groups. As such, any difference observed is more attributable to year to year changes in ecological variables (such as land cover characteristics) or occurrence itself. The first stage of calculating our test statistic is to train a random forest on both $\mathcal{D}_{08-09}$ and $\mathcal{D}^{NN}_{10-13}$. We then use these forests to make predictions at fixed test points, from which several summary statistics are calculated. 

Our test set consists of $166\times1000$ points, with $1000$ points taken for each day in the fall. We construct this test set by sampling 1000 locations from a $3\text{km} \times 3\text{km}$ grid covering the east coast, referenced with their land cover and elevation characteristics, as well as  \texttt{max\_temp} for that day. The maximum temperature information included is the expected daily maximum temperature, estimated from the 1980-2007 temperature information provided by DayMet.  The variables associated with the eBird user (e.g. \texttt{eff\_dist}, \texttt{eff\_hours}, and \texttt{n\_obs}) are set to 1 uniformly, to again represent typical eBird user levels.

Let $R_{08-09}(\cdot)$ denote the prediction function of the RF trained on $\mathcal{D}_{08-09}$.  Then $f_{08-09}$ is defined as
\[
f_{08-09}(t) := \frac{1}{1000}\sum_{k=1}^{1000}R_{08-09}(\bm{x}_{k,t}), \quad t \in \{200,...,365\}
\]

\noindent where $\bm{x}_{k,t}$ is a point in the test set corresponding to time $t$. Thus, since the test points are stratified by time, $f_{08-09}(t)$ denotes the average over predictions made at all 1000 test points on each day and therefore represents a time-averaged version of the raw RF prediction function.  The function $f_{10-13}(t)$ is defined in exactly the same fashion for a RF trained on $\mathcal{D}^{NN}_{10-13}$. 

Recall that the original hypothesis was that Tree Swallows remained in BCR30 longer in 2008-2009 than in 2010-2013, followed by a sharp decline in numbers. Preliminary analysis in Figure \ref{RF_TSO_PE} supports this hypothesis, but we now want to evaluate the statistical significance of the evidence. Formally, in early fall (\texttt{DoY} 200-264), it appears that $f_{08-09} > f_{10-13}$.  Later in the fall (\texttt{DoY} 265-310), it appears that $f_{08-09} < f_{10-13}$ and finally as winter sets in (\texttt{DoY} 311-365), we see $f_{08-09} \approx f_{10-13}$. We therefore partition our time frame into three disjoint time periods, $T_1 = \{200,...,264\}$, $T_2 = \{265,...,310\}$, and $T_3 = \{311, ...,365\}$, and let $I_{T_i}(t)$ be an indicator function for each period. We then consider the restricted functional observations
\[
f^{(i)}_{08-09}(t) := f_{08-09}(t)I_{T_i}(t) \ \text{ for }  i = 1,2,3
\]

\noindent which are defined in the same fashion for $f^{(i)}_{10-13}(t)$. Each of these restricted functional observations is then incorporated into test statistics to evaluate following sets of hypotheses 
\begin{equation} \label{H0_D1}
\begin{split}
\text{H}_{0,i}:& \ D^{(i)}_{08-09}(\bm{X}, Y) =D^{(i)}_{10-13}(\bm{X}, Y)  \\
\text{H}_{1,i}:& \ D^{(i)}_{08-09}(\bm{X}, Y) \neq D^{(i)}_{10-13}(\bm{X}, Y).
\end{split}
\end{equation}

To evaluate these hypotheses, we begin by calculating the prediction functions over time using the original datasets and then, for each of many iterations, we permute the \texttt{year} covariate, re-partition the data into the two groups consisting of data from 2008-2009 and 2010-2013, and construct the new RF prediction functions. 

Permutation tests for hypotheses of this form reject $H_0$ if the test statistic, $T_0$, calculated on the original data, falls in the extreme (upper or lower $\alpha/2$) quantile of the permutation distribution of test statistics. Formally, given two sets of data $\{X_i\}_{i=1}^n$ and $\{Y_i\}_{i=1}^m$, let $\mathcal{G}$ be the group of all permutations of the indices $1,..., m+n$. Then, consider a statistic of the form $T = T(Z_1,... Z_{m+n})$, and let $T_0$ be the statistic calculated on the original data. A p-value for the hypothesis the null hypothesis  $H_0: D(X) =  D(Y)$ is given by
\[
p = \frac{1}{|\mathcal{G}|}\sum_{\pi \in \mathcal{G}} I(|T_0| > |T(Z_{\pi(1)}, ..., Z_{\pi(m+n)})|) .
\] 
Note that $|\mathcal{G}| = \binom{m+n}{n}$ which is quite large, so we instead sample 1000 draws from the permutation distribution uniformly at random, which maintains the size of the test at $\alpha$ \citep{Lehmann2006}. Permutation tests offer flexibility in the choice of test statistic, and different test statistics offer different levels of power. As such, we consider the following two measures of functional distance
\begin{equation} \label{eqn:teststats}
\begin{split}
KS &= \sup_{t\in T_1\cup T_2\cup T_3} \left| f_{08-09}(t) - f_{10-13}(t) \right|  \\
\Delta_i &= \frac{1}{|T_i|}\sum_{t\in T_i} (f^{(i)}_{08-09}(t) - f^{(i)}_{10-13}(t)), \quad i = 1,2,3 .
\end{split}
\end{equation}
The first measure in \eqref{eqn:teststats} refers to the Kolmorgorov-Smirnov statistic, traditionally used to test hypotheses about distribution functions.  The use of this statistic for two sample functional testing procedures was studied by \citet{hall2007two}.  $KS$ is calculated across the full time period, and then used for testing for an overall difference in the underlying distributions. 
In contrast, our raw distance measures $\Delta_1, \Delta_2, \text{ and } \Delta_3$ are designed to test for equality of the underlying distributions only in time periods $T_1, T_2, \text{ and } T_3$, respectively.  Based on the visual evidence in Figure \ref{RF_TSO_PE}, we may expect to see a difference during the first two time periods, but likely not during the third time period.

\subsubsection{A Computationally Efficient Alternative Testing Procedure}
\label{sec:func_data}

The procedure described above maintains many of the desirable statistical properties of permutation tests, such as exactness under any distribution. As such, we refer to it as the \textit{canonical} permutation test. However, permutation tests were developed for situations where test statistics are easily calculated. Because our test statistic involves training of two random forests, there is substantial computational burden incurred in conducting each test. Indeed, running the full test requires constructing  $2 \times N_{\text{Perm}} \times B$ decision trees, where $B$ is the size of each forest.  As such, we now propose a computationally efficient alternative.

Random forest predictions can be written as a function of the training data, the test point, and a collection of randomization parameters, $\bm{\xi} = \{\xi_1, ..., \xi_B\}$, which dictate the feature subsetting and resampling used in each tree. For a given test point $\bm{x}$, the random forest prediction $R(\bm{x} ; \mathcal{D}, \bm{\xi})$ can be written as
\[
R(\bm{x} ; \mathcal{D}, \bm{\xi}) = \frac{1}{B}\sum_{k=1}^B T(\bm{x}; \mathcal{D}, \xi_k)
\]
where $T(\cdot ; \mathcal{D}, \xi)$ is a standard CART decision tree trained on $\mathcal{D}$ using randomization $\xi$. In a random forest, the randomization parameters are drawn in an iid fashion, so that for any point $\bm{x}$ and any number of trees $B$, $\{T(\bm{x}; \mathcal{D}, \xi_k)\}_{k=1}^B$ is an iid sequence conditional on $\mathcal{D}$. Now, we can appeal to the classical De Finetti's Theorem \citep{DeFinetti1937} for infinitely exchangeable random variables, which states that \textit{a sequence of infinitely exchangeable random variables is exchangeable if and only if it is iid conditional on some other random variable.} The sequence of trees used in a random forest are iid, conditional on the data, and therefore are infinitely exchangeable. Moreover, suppose we partition a collection of $B$ trees into $k$ subgroups, each consisting of $B/k$ trees, and form $k$ random forests from these trees. Then, the same argument gives that $R_1(\bm{x} ; \mathcal{D}), ..., R_k(\bm{x} ; \mathcal{D})$ is infinitely exchangeable, and further that the functional observations (like those used in the test statistics in \eqref{eqn:teststats}) are realizations of an infinitely exchangeable sequence of functions. As such, if we train $B$ trees, and then randomly stratify the trees into $k$ forests of equal size, we have an exchangeable sequence of functions. 

Exchangeability is fundamental to the exactness of permutation tests. In fact, a permutation test is fundamentally a test of \textit{exchangeability} - for two groups of data $X_i \iid P$ and independently, $Y_i \iid Q$, the data are exchangeable if and only if $P \equiv Q$. To see this, note that under an exchangeability assumption:
\begin{equation} \label{exchgcond}
D_0 := D(X_1,..., X_n, Y_1,..., Y_m) = D(Z_{\pi(1)},..., Z_{\pi(m+n)}) := D_{\bm{Z}_\pi} \quad \forall \pi
\end{equation}
where $\bm{Z}_\pi$ is any permutation of the $X_i$ and $Y_i$. Thus, for any given test statistic (i.e.\ a function of the $m+n$ observations), the quantile of the observed test statistic across all possible permutations should approximately follow a uniform distribution \citep{Pesarin2010}. For iid observations, \eqref{exchgcond} factors as
\begin{equation*}
D(X_1,..., X_n, Y_1,..., Y_m) = \prod_{i=1}^n P(X_i) \times \prod_{i=1}^m Q(Y_i) \stackrel{\text{exchangeable}}{\equiv} \prod_{i=1}^{n+m} P(Z_\pi(i)).
\end{equation*}
Thus for iid data, the finite sample permutation test provides an exact test for hypotheses about equality of distribution. Indeed, as a result of \eqref{exchgcond}, for any statistic $T(\cdot)$, $T(\bm{Z}) \stackrel{d}{=} T(\bm{Z}_\pi)$. A more rigorous argument for the validity of the tests is presented in \citet{Lehmann2006}.

In the set up described, we sample 20 exchangeable functions, $\{f_{k, (08-09)}\}_{k=1}^{20}$ and $\{f_{k, (10-13)}\}_{k=1}^{20}$ each using $50$ identically trained decision trees. Treating the observed functions $f_{08-09}$ and $f_{10-13}$ as observations from functional distributions $\mathcal{F}_{08-09}$ and $\mathcal{F}_{10-13}$,  our goal is to determine whether these distributions that generated our observed prediction functions are, in fact, the same. More explicitly, we consider hypotheses of the form
\begin{equation} \label{H0_func}
\begin{split}
\text{H}^f_0:& \ \mathcal{F}_{08-09} = \mathcal{F}_{10-13}  \\
\text{H}^f_1:& \ \mathcal{F}_{08-09} \neq \mathcal{F}_{10-13} .
\end{split}
\end{equation}
\noindent where the $H^f$ notation is to distinguish these hypotheses from those in \eqref{H0_dist}. It should be noted that even under $H^f_0$, the forests are not exactly exchangeable between groups since the conditioning random variable (the datasets, $\mathcal{D}_{08-09}, \mathcal{D}^{NN}_{10-13}$) are different. As such, there will be stronger dependence within the groups of trees. To ameliorate this, we impose an additional condition on the construction of the random forests. In particular, instead of bootstrapping, we now subsample observations, i.e.\ each tree is trained on $k < n$ observations, without replacement. We use a dynamic subsampling rate, with $k_n = n^p$ for some $p \in (0,0.5)$, so that $\lim_{n\to\infty} k_n/n = 0$. This ensures that the decision trees used are asymptotically independent, which means that the dependence between tree predictions dies off as $n\to\infty$. Thus, the within group and between group dependences approach each other as $n\to\infty$. We note that this is a standard requirement imposed upon random forest construction in the random forest theory, such as in \citet{mentch2016quantifying, Wager2017, scornet2015consistency}. The choice of mini-ensembles of size 50 is done to balance predictive accuracy and the higher within sample dependence.

While $H^f_0$ does not imply $H_0$, rejecting $H_0^f$ supports the notion that migration patterns in the years 2008 and 2009 differed significantly from those observed from 2010-2013. Note that because a random forest is simply an average of decision trees, we can reformulate each of the statistics in \eqref{eqn:teststats} by substituting in $f_{(08-09)}(t) = \frac{1}{20} \sum_{k=1}^{20} f_{k,(08-09)}(t)$ and likewise for $f_{(10-13)}(t)$. Then we shuffle the functional observations between the 2008-2009 group and the 2010-2013 group many times, at each stage calculating the statistics in \eqref{eqn:teststats}. That is, to form a permuted random forest, we permute groups of decision trees rather than the data itself, so that we now only have to train $2B$ trees. Similarly, for the temporally segmented test, each restricted functional observation is from some distribution $\mathcal{F}^{(i)}_{08-09}$ or $\mathcal{F}^{(i)}_{10-13}$, leading naturally to hypotheses of the form
\begin{equation} \label{H0_func_i}
\begin{split}
\text{H}^f_{0, i}:& \ \mathcal{F}^{(i)}_{08-09} = \mathcal{F}^{(i)}_{10-13}  \\
\text{H}^f_{1,i}:& \ \mathcal{F}^{(i)}_{08-09} \neq \mathcal{F}^{(i)}_{10-13} .
\end{split}
\end{equation}
We take advantage of the bagging structure inherent to random forests, but the same framework, which we refer to as the \textit{functional} permutation test, could be applied to any bagged learner. 

\subsection{Global Test Results}
\label{subsec:globtest_r}

\begin{figure}
    \centering
    \begin{subfigure}[t]{0.5\textwidth}
        \centering
        \includegraphics[width = 19pc]{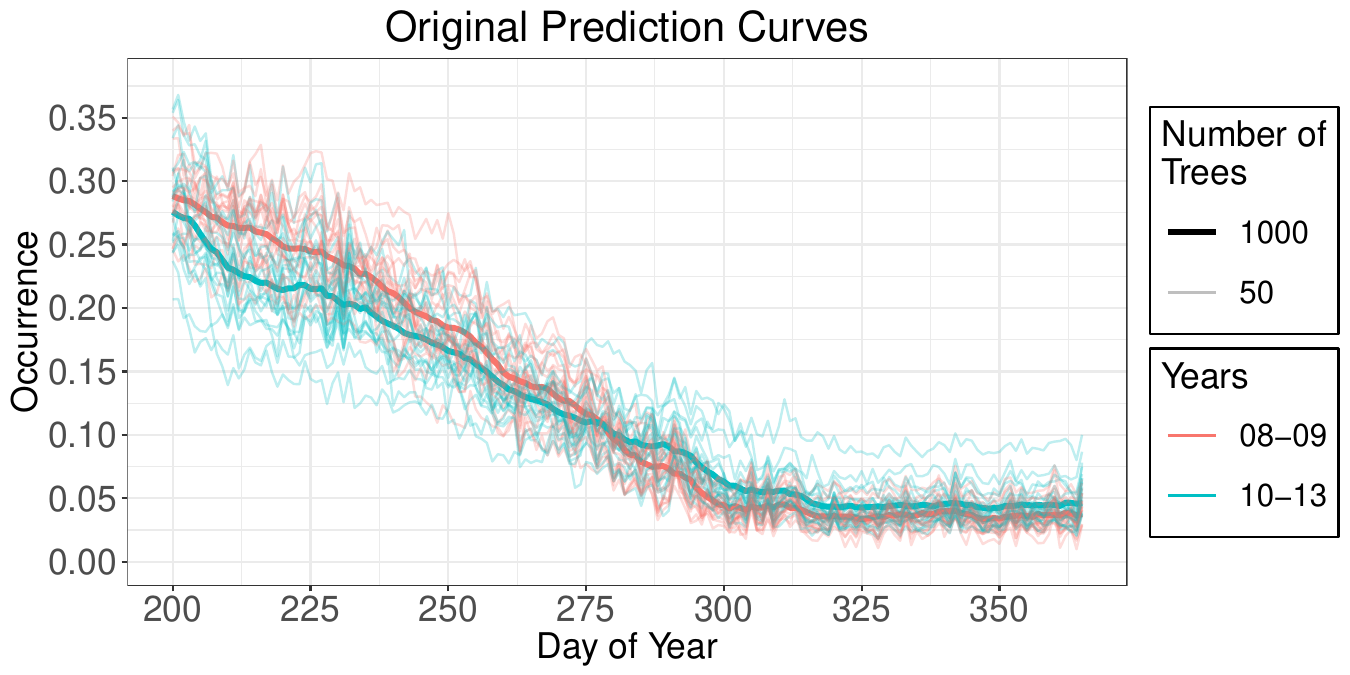}
        \caption{$f_{08-09}, f_{10-13}$, functional data unpermuted}
    \end{subfigure}%
    \begin{subfigure}[t]{0.5\textwidth}
        \centering
        \includegraphics[width = 19pc]{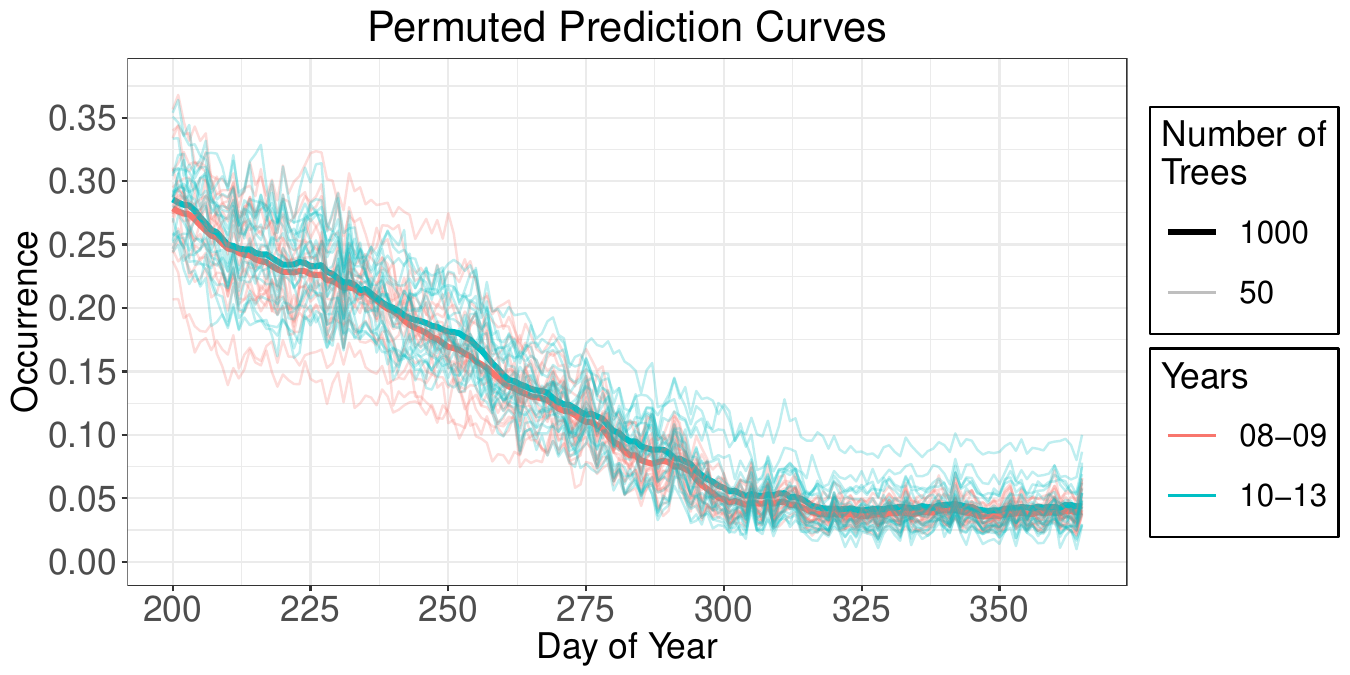}
        \caption{$f_{08-09}, f_{10-13}$, functional data permuted}
    \end{subfigure}
    \caption{Prediction curves generated by RFs which use all covariates, including \texttt{max\_temp} and \texttt{DoY}. These are used in the functional permutation test. Lighter lines show the collection of functional data, darker lines show average that forms the full RF function.}
    \label{Full_Curves}
\end{figure}
To implement the canonical permutation test, we utilize the \texttt{randomForest} package in \texttt{R} to calculate the RF predictions \citep{Liaw2002}.  As in Section \ref{subsec:RFheuristic}, the RFs are trained on the entire set of predictors, including both \texttt{DoY} and \texttt{max\_temp}. To account for the correlation between \texttt{max\_temp} and \texttt{DoY}, we conduct two additional followup versions of the original permutation tests:  one with \texttt{DoY} included and \texttt{max\_temp} removed and one with \texttt{max\_temp} included and \texttt{DoY} removed. The functional permutation test is conducted using 20 functional observations in each group, using a subsampling rate of $a_n = n^{0.55}$ and setting $\texttt{mtry} = 7$. These constraints on the tree construction worsens the predictions of the individual models, but further weakens the dependence between the functional data. 

The p-values obtained from the canonical permutation test are shown in the second row of Table \ref{tab:tabthree}.  Based on these results alone, there does not appear to be strong evidence of a difference in the underlying functional distributions, even early in the migration period. However, a more compelling story appears in the results of functional permutation test. The associated functions, $\{f_{k (08,09)}\}_{k=1}^{20}$ and $\{f_{k, (10-13)}\}_{k=1}^{20}$, along with their averages, are shown in Figure \ref{Full_Curves}, along with an example of a permutation of the functional data. Based on a visual inspection, it appears that for RFs trained on the original 2008-2009 data ($\mathcal{D}_{08-09}$) and the reduced nearest neighbor 2010-2013 data ($\mathcal{D}_{10-13}^{NN}$), $f_{08-09} > f_{10-13}$ until around \texttt{DoY} 280, with negligible differences thereafter. The p-values from the functional permutation test are presented in Table \ref{tab:tabthree}, and provide strong evidence for a difference in migration patterns. In particular, we are able to reject $H_0^f$, for the full feature and \texttt{max\_temp} models at any reasonable level $\alpha$.

\begin{figure}[t]
    \centering
    \begin{subfigure}[t]{0.5\textwidth}
        \centering
        \includegraphics[width = 19pc]{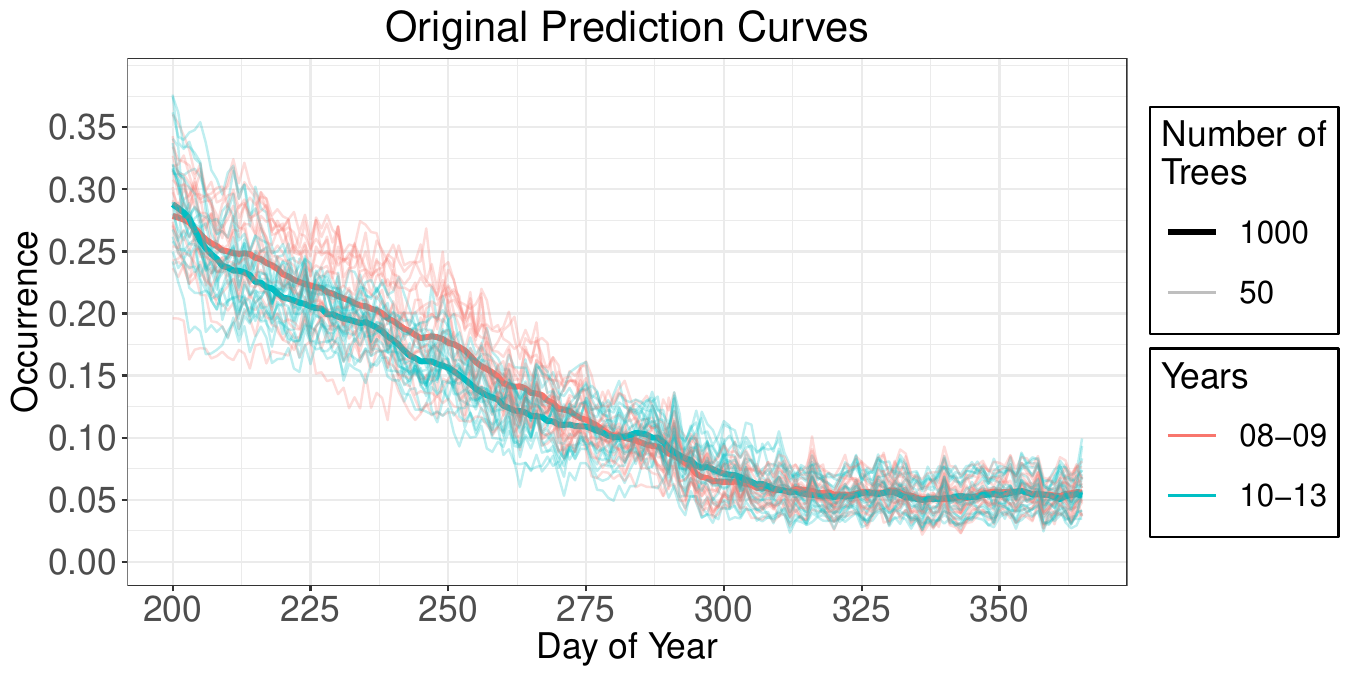}
        \caption{Original data, \texttt{max\_temp} removed}
    \end{subfigure}%
    \begin{subfigure}[t]{0.5\textwidth}
        \centering
        \includegraphics[width = 19pc]{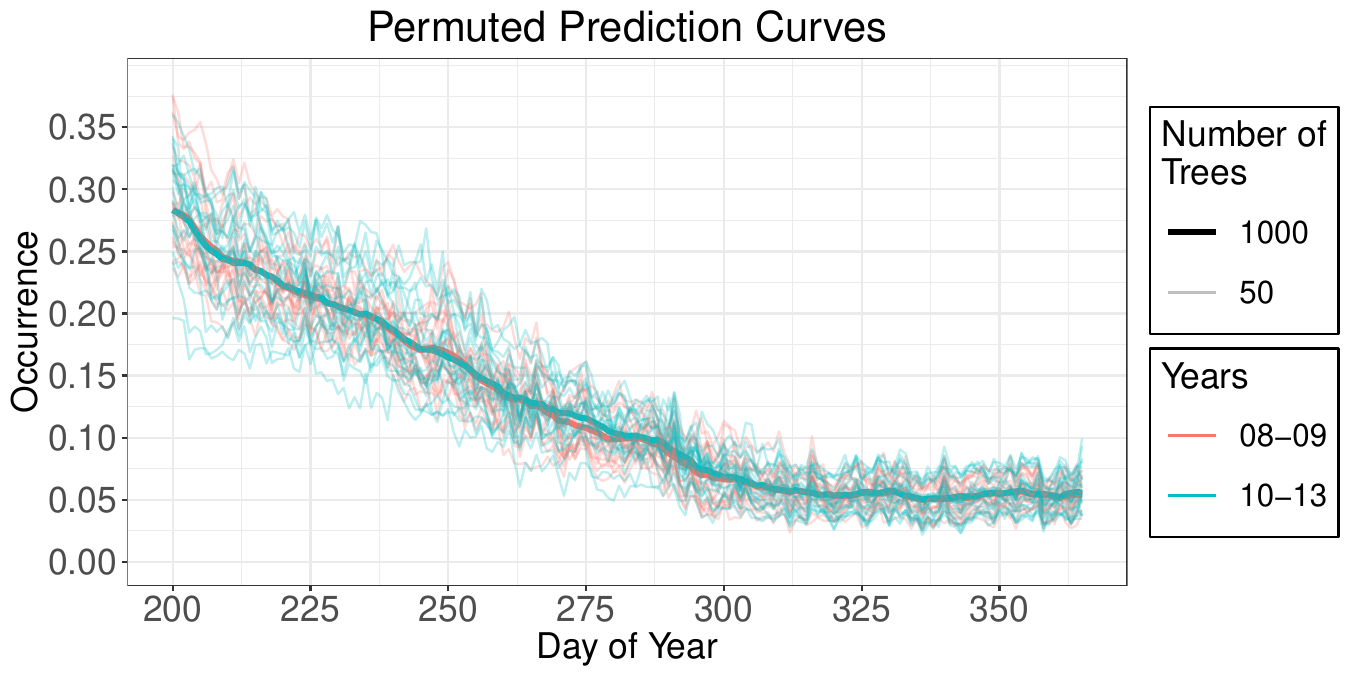}
        \caption{Permuted data, \texttt{max\_temp} removed}
    \end{subfigure} 
    \begin{subfigure}[t]{0.5\textwidth}
        \centering
        \includegraphics[width = 19pc]{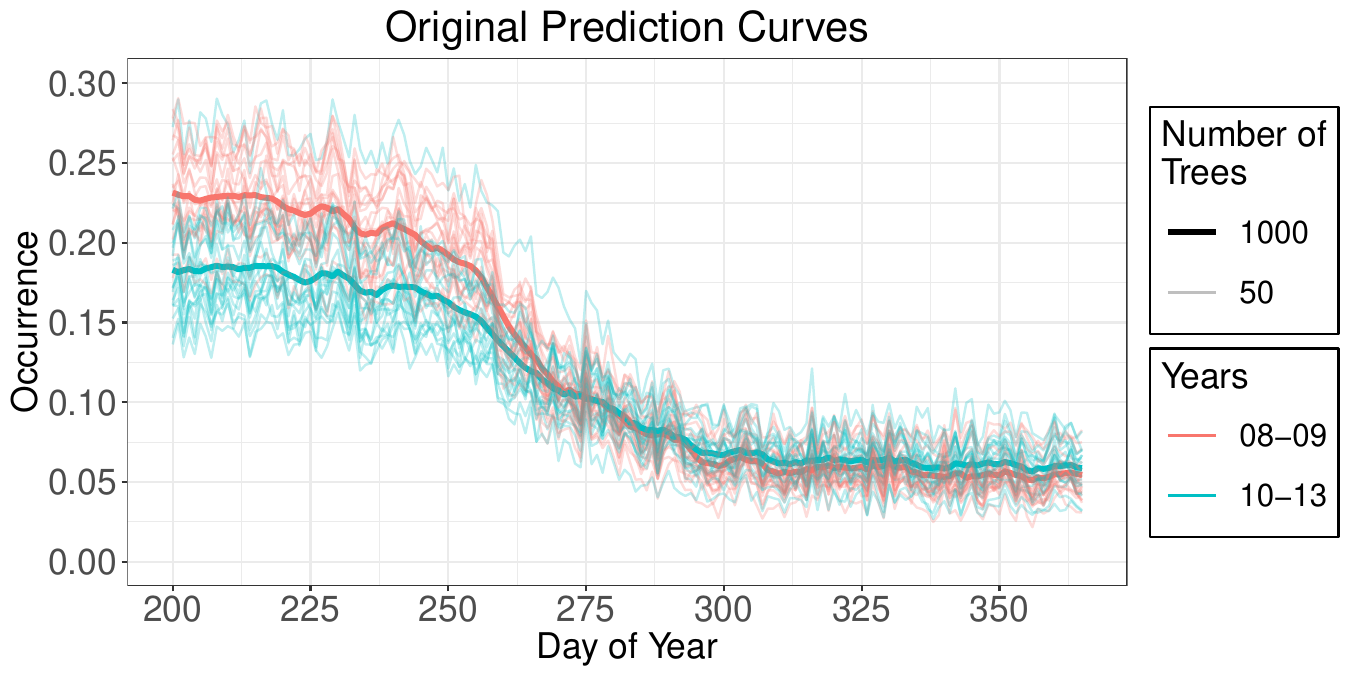}
        \caption{Original data, \texttt{DoY} removed}
    \end{subfigure}%
    \begin{subfigure}[t]{0.5\textwidth}
        \centering
        \includegraphics[width = 19pc]{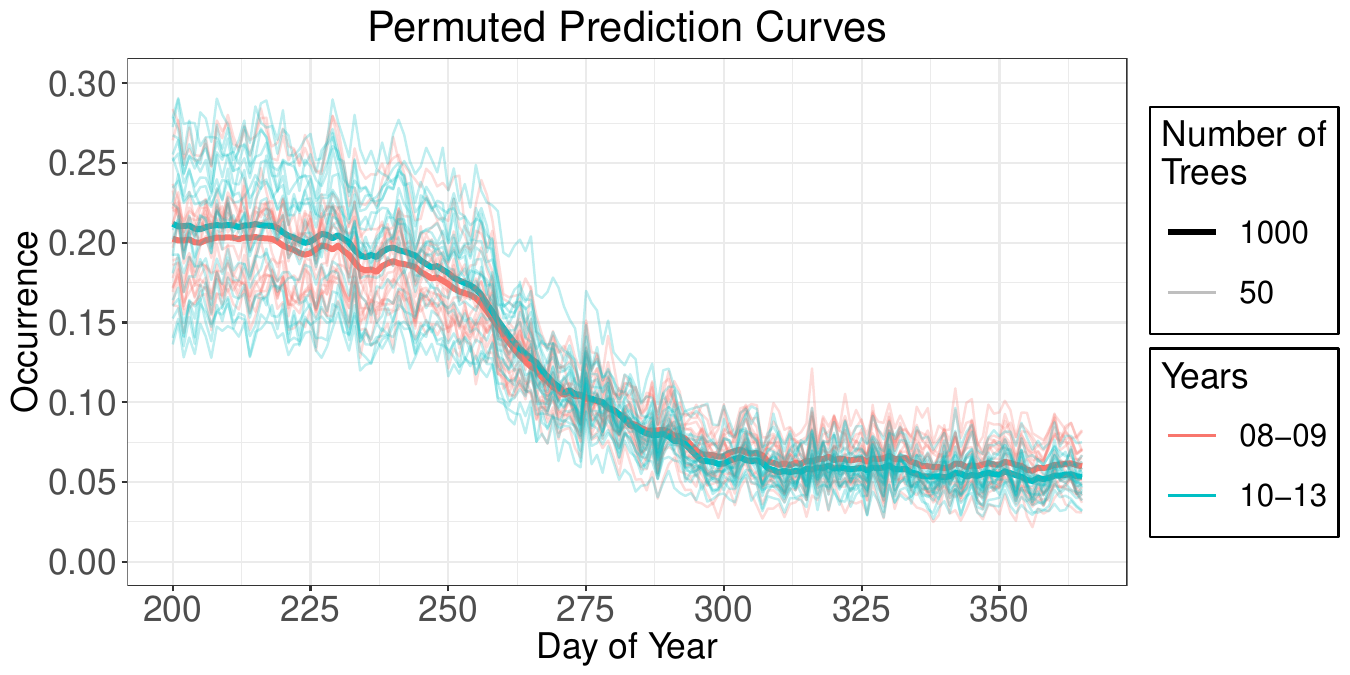}
        \caption{Permuted data, \texttt{DoY} removed}
    \end{subfigure}
    \caption{Prediction curves generated by RFs with \texttt{DoY} included and \texttt{max\_temp} removed (top row, (a) and (b)) and with \texttt{max\_temp} included and \texttt{DoY} removed (bottom row, (c) and (d)).  Lighter lines show collection of functional data, darker lines show average that forms the full RF function.}
    \label{PermCurves}
\end{figure}

The RF prediction functions corresponding to the \texttt{max\_temp} only and \texttt{DoY} only models, with both the original and (randomly selected) permuted datasets, are shown in Figure \ref{PermCurves}.  Here we begin to see evidence for the importance of \texttt{max\_temp}:  when only \texttt{DoY} is used as a predictor, the prediction functions trained on the original datasets closely resemble those trained on the permuted data.  However, when \texttt{max\_temp} is included and \texttt{DoY} is removed, we see a clear difference in predicted occurrence until midway into the migration season, a story which closely matches the anecdotal accounts from the ornithological community. Moreover, the raw statistic values in Table \ref{tab:stat_tab} show that the greatest differences (however those differences are measured) are consistently observed in the \texttt{max\_temp} model. 

\begin{table}[ht]
\centering
\begin{tabular}{|c|cccc|}
\hline
Test Statistic &  $KS$ & $\Delta_1$ & $\Delta_2$ & $\Delta_3$ \\ 
\hline
\texttt{DoY} \& \texttt{max\_temp} included  & 0.05112 & 0.02419 &\textbf{ -0.00741} & \textbf{-0.00816} \\ \hline
\texttt{DoY} included; \texttt{max\_temp} removed  & 0.02718 & 0.01530 & 0.00037 & 0.00096 \\ \hline
\texttt{max\_temp} included; \texttt{DoY} removed & \textbf{0.05329} & \textbf{0.03651} & -0.00170 & -0.00533 \\ 
\hline
\end{tabular}
\caption{Observed values for each of the test statistics in \eqref{eqn:teststats}. Bolded values are the largest magnitude differences for each test.} \label{tab:stat_tab}
\end{table}

\begin{table}
\centering
\begin{tabular}{|c||cccc|cccc|}
  \hline
Test Statistic & $KS$ & $\Delta_1$ & $\Delta_2$ & $\Delta_3$ & $KS$ & $\Delta_1$ & $\Delta_2$ & $\Delta_3$ \\ 
\hline
Null Hypothesis Tested  & $\text{H}_0$ & $\text{H}_{0,1}$ & $\text{H}_{0,2}$ & $\text{H}_{0,3}$  & $\text{H}^f_0$ & $\text{H}^f_{0,1}$ & $\text{H}^f_{0,2}$ & $\text{H}^f_{0,3}$  \\
\hline
\texttt{DoY} \& \texttt{max\_temp} included  & 0.196 & 0.075 & 0.768 & 0.829 & 0.002 & 0.007 & 0.103 & 0.954 \\ 
\hline
  \texttt{DoY} included; \texttt{max\_temp} removed & 0.122 & 0.034 & 0.544 & 0.163  & 0.565 & 0.182 & 0.676 & 0.677 \\ 
\hline
\texttt{max\_temp} included; \texttt{DoY} removed  & 0.001 & 0.000 & 0.299 & 0.775 & 0.070 & 0.055 & 0.020 & 0.711 \\ 
\hline
\end{tabular}
\caption{P-values for the canonical permutation test (left) and functional test (right); Tests are done with all covariates included in the datasets (row three), \texttt{DoY} included and \texttt{max\_temp} removed (row four), and \texttt{max\_temp} included and \texttt{DoY} removed (row five).} \label{tab:tabthree}
\end{table}

The p-values resulting from the followup tests appear to tell a similar story, in both the functional and canonical permutation test.  From Table \ref{tab:tabthree} we see that when \texttt{max\_temp} is removed, the smallest p-value is only 0.182 corresponding to the test for raw differences in time period one as measured with $\Delta_1$.  However, when \texttt{max\_temp} is included and \texttt{DoY} removed, the largest p-value among the first four tests is only 0.07.  The p-value from the final test for raw differences in the third time period (measured by $\Delta_3$) is large at 0.711, but recall that this is what was expected as the prediction curves appear very similar in all cases late in the season. A similar pattern appears in the p-values in Table \ref{tab:tabthree}. 


Before continuing with the localized tests the in the following section, we acknowledge that the p-values from the canonical permutation tests, though reasonably small and substantially lower than in the other tests, fail to surpass the commonly accepted $\alpha=0.05$ threshold in all but one instance and too large to conclude that migration patterns differed significantly in the two sets of years.  However, note that these tests are conservative in two ways.  First and most obviously, permutation tests themselves suffer lower power than their parametric counterparts.  More subtly, the RF prediction curves generated by the data from 2010-2013 were not trained on the full available dataset, but were trained on a carefully selected subset $\mathcal{D}_{10-13}^{NN}$ designed to spatiotemporally mimic the observations collected in 2008-2009.  Given this, it is reasonable to interpret the results of the canonical test as providing at least moderate evidence for a difference in migration patterns that is influenced by \texttt{max\_temp}. The same patterns appear in the functional test results, in greater magnitude, providing stronger evidence of a yearly difference in occurrence patterns. The localized tests in the following section allow for a more direct means of measuring the precise questions of interest and provide more decisive evidence.

\section{Localized Hypothesis Testing}
\label{sec:loctest}
The tests carried out in the previous section average out occurrence and the covariate influence across the entire BCR30 spatial region, neglecting the local factors that likely play a role in Tree Swallow migration. Here we evaluate the predictive utility of \texttt{max\_temp} at a local spatial level by reexamining the above hypotheses at small sets of specific test points that are local with respect to both space and time.  In order to conduct these tests, we make use of recent asymptotic theory which provides a closed form distribution for predictions originating from subsampled ensemble estimators.  We summarize these recent theoretical advances in the following section before presenting our testing procedure.


\subsection{Random Forests as U-Statistics}
\label{subsec:ssrf}

Our localized tests rely on the work of \citet{mentch2016quantifying} from which we now briefly review some key results.  Suppose we have a training sample $\mathcal{D} = \{ Z_1, ..., Z_n\}$ consisting of $n$ iid observations from some distribution $F_Z$ with which we construct a (possibly randomized) ensemble consisting of $m$ base learners, each built with a subsample of size $k$, and use this ensemble to predict at some location $\bm{x}$.  Denote each base learner by $h$ so that we can write the expected prediction as $\theta_k = \theta_k(\bm{x}) = \mathbb{E} h(\bm{x}; Z_1, ..., Z_k)$ and the (empirical) ensemble prediction as $\hat{\theta}_k = \hat{\theta}_k(\bm{x}) = \frac{1}{m} \sum_{i=1}^{n} h(\bm{x}; Z_{1}^{*}, ..., Z_{k}^{*})$ where each collection $(Z_{1}^{*}, ..., Z_{k}^{*})$ represents a subsample of size $k$ from $\mathcal{D}$. Then, under some regularity conditions introduced in \citet{mentch2016quantifying} later weakened in \citet{Peng2019},  
\begin{equation}
\frac{\sqrt{m} (\hat{\theta}_{k} - \theta_k)}{\sqrt{\frac{k^2}{\alpha} \zeta_1 + \zeta_k }} \overset{d}{\rightarrow} \mathcal{N}(0,1)
\label{eqn:RFdist}
\end{equation}
 
\noindent where $\alpha = \lim_{n \rightarrow \infty} n/m$ and the other variance parameters are of the form
\begin{equation}
\zeta_c = \text{cov}(h(Z_1, ..., Z_c,Z_{c+1}, ..., Z_{k}), h(Z_1, ..., Z_c,Z_{c+1}^{'}, ..., Z_{k}^{'}))
\label{eqn:RFvar}
\end{equation}

\noindent for $1 \leq c \leq k$ and where $Z_{c+1}^{'}, ..., Z_{k}^{'}$ denote additional iid observations from $F_Z$.  

Importantly, this result can be utilized to construct formal hypothesis tests of variable importance.  Suppose we have $p$ covariates $X_1, ..., X_p$ and we want to test the predictive importance (significance) of $X_1$.  Let $\hat{d}_{i} = R(\bm{x}_{i}) - R_{\pi_1}(\bm{x}_{i})$ denote the difference in predictions between a forest trained on $\mathcal{D}$ and another trained on $\mathcal{D}_{\pi_1}$ in which $X_1$ is permuted. 
Consider $N$ such prediction points with differences denoted by $\hat{d} = (\hat{d}_1, ..., \hat{d}_N)^T$.  \citet{mentch2016quantifying} show that when the same subsamples are used to construct the trees in each random forest, the differences are infinite-order U-statistics and thus follow the asymptotic distribution in \eqref{eqn:RFdist}. Let $\hat{\Sigma}_d$ be the estimated covariance matrix of the $\hat{d}_i$. Then, given our vector of pointwise differences, $\hat{d}^T  \, \hat{\Sigma}_d^{-1} \, \hat{d} \sim \chi_{N}^{2}$ and we can use this as a test statistic to formally evaluate the hypotheses
\begin{equation} \label{HypMH}
\begin{split}
\text{H}_0:& \quad \E R(\bm{x_i}) = \E R_{\pi_1}(\bm{x_i}) \quad \text{ for all } i \in \{1, ..., N\} \\ 
\text{H}_1:& \quad \E R(\bm{x}) \neq \E R_{\pi_1}(\bm{x}) \quad \text{ for some } i \in \{1, ..., N\} 
\end{split}
\end{equation}
\noindent where the expectation is taken with respect to the training data and randomization.  This procedure naturally extends to the more general case where any subset of the features is tested for significance by simply permuting that entire subset of features.  Furthermore, this procedure remains valid whenever those features are simply removed from the alternative random forest instead of being permuted, though the permutation-based approach is generally considered more robust and reliable \citep{mentch2016quantifying}.

\subsection{Local Influence of Maximum Temperature}
\label{subsec:localMaxTemp}

We return now to the question of determining whether maximum daily temperature can partially explain the different Tree Swallow patterns of occurrence observed in 2008-2009.   The global tests in the previous section suggested that \texttt{max\_temp} may provide information about the interannual variation in occurrence beyond what is provided by seasonal effects alone captured by \texttt{DoY}. We therefore want to distill the predictive influence of \texttt{max\_temp} from that of \texttt{DoY}. In this section, we consider testing for the variable importance of \texttt{max\_temp\_anomaly}. 

To fit this into the hypothesis testing framework described above, we calculate one subsampled random forest with the original data and another with a permuted version of \texttt{max\_temp\_anomaly}.  Note that because the hypotheses in \eqref{HypMH} are evaluated at only fixed test points, careful selection of these points is important. Since we are interested in evaluating the hypotheses across a variety of locations and times, we stratify our test points by location and conduct 6 different tests. 

\begin{figure}[t]
\begin{floatrow}
\ffigbox{%
     \includegraphics[width = .45\textwidth]{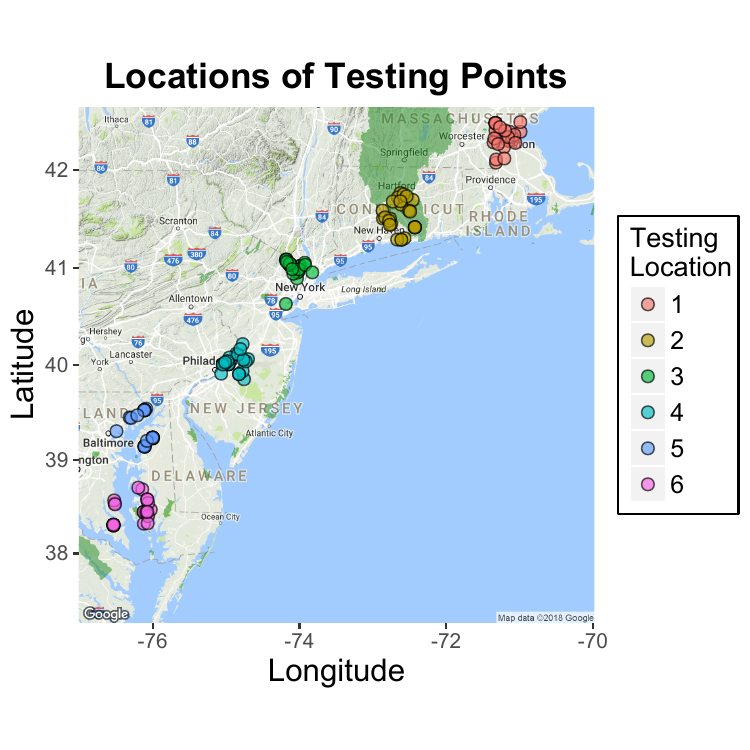}
     }{
\caption{Test points selected for study. Light green overlay indicates Fish and Wildlife Service wildlife refuges.}
\label{TestingMap}
}
\capbtabbox{%
\begin{tabular}{cccc}
  \hline
\textbf{Testing Location} & \textbf{Test Statistic} & \textbf{p-value}  \\
  \hline
1  & 38.07 & 0.04553 \\ 
  2  & 58.12 & 1.889E-04 \\ 
  3  & 58.21 & 1.835E-04 \\ 
4& 62.14 & 5.275E-05 \\ 
5  & 59.93 & 1.068E-04 \\ 
6  & 28.44 & 0.2880 \\ 
   \hline
\end{tabular}

}{%
 \caption{Test statistics and p-values for the hypotheses in \eqref{HypMH} at the points show in Figure \ref{TestingMap}. } \label{tab:tabfour}}
\end{floatrow}
\end{figure}

The training and test set used here are selected from points inside wildlife refuge areas. Wildlife refuges are of particular interest because they include areas that are resistant to local environmental changes due to the environmental protections in place, helping to isolate the predictive influence of regional temperature fluctuations on Tree Swallow occurrence.  In total, we select 6 groups of 25 test points each, which are subsequently removed from the training set. These 6 groups and points are shown in Figure \ref{TestingMap}. Spatial centers for each of these regions were selected based on a high density of observations and the test points were selected uniformly at random from within a 0.3 decimal degree radius.  The final training set consists of 25727 observations.

We apply the above hypothesis testing procedure at each of our 6 regional test locations, building separate ensembles for each location.  As in Section \ref{sec:glob}, we make all features available for splitting at each node in each tree so that our random forest procedure reduces to subsampled bagging (subbagging). These tests are implemented using the \texttt{rpart} package in \texttt{R} to construct the regression trees \citep{Thernau2017}.  Keeping with the recommendations of \citet{mentch2016quantifying}, we take our subsample size to be $k=160 \approx \sqrt{25727}$; in general, larger subsamples can be used if base learners are constructed in an alternative fashion to comply with honesty and regularity conditions \citep{Wager2017}. We build $1.25 \times 10^7$ trees for each ensemble, to attain high precision in the estimation of the covariance matrix, $\hat{\Sigma}_d$. 
Table \ref{tab:tabfour} summarizes the test statistics and p-values obtained from the tests in each region.  These local tests for the significance of \texttt{max\_temp\_anomaly} suggest that the anomaly is predictive of occurrence in testing locations 2-5, with less significance in location 1 and no significance in location 6. These suggest a transition zone within BCR30 between testing locations 1 \& 6 where \texttt{max\_temp\_anomaly} is important in predicting occurrence. North of this zone, temperatures may be too cold to allow insect activity in the fall and south of this zone, temperatures may be warm enough to allow insect activity year round. 
\section{Discussion}
\label{sec:conclusions}
\subsection{Ornithological Implications}
Our goal in this work was to thoroughly examine Tree Swallow migration patterns from recent years and to examine whether temperature changes are predictive of migration differences across years.  The global hypothesis tests evaluated over the entire BCR30 region in Section \ref{sec:glob} provided evidence for the hypothesis that the seasonal patterns of distributions indeed differed in the years 2008 and 2009.  The fact that this difference no longer seemed apparent whenever \texttt{max\_temp} was excluded as a predictor supports the hypothesis that temperature can partially explain year-to-year variation in occurrence.  The corresponding localized hypothesis tests carried out at specific locations along the Tree Swallow migration route in Section \ref{sec:loctest} provide formal justification for the importance of maximum temperature beyond being merely a correlate of some other seasonally varying effect. These results are especially important in the context of climate change, demonstrating that variation in ambient temperatures is predictive of the mortality and migration of a wild bird, supporting conclusions of other ecological work \citet{LaSorte2016}. 
\subsection{Methodological Discussion}
We present two forms of black box inference: development of non-parametric permutation-style tests that are agnostic to the underlying procedure, and asymptotic results about the predictions of ensemble learners. The recent asymptotic results for infinite-order U-statistics allowed us to model the data through a series of flexible but complex black-box models -- in our case, regression trees -- while retaining the ability to formally characterize results. We also present a classical statistical argument for the validity of the functional permutation test. 
This procedure maintains asymptotic validity (in terms of controlling Type I error), and requires training substantially fewer trees than  the canonical permutation test. We also note that this procedure, despite being non-parametric, requires far fewer trees than even the theoretical results of \citet{mentch2016quantifying} and \citet{Wager2017}, making it a much more practical tool for applications like our case study.

\bigskip
\begin{center}
{\large\bf SUPPLEMENTARY MATERIAL} \\
\end{center}
A supplementary file containing details about some of the results is provided.  \texttt{R} files are provided for managing the data, performing the analysis, and generating the figures. 

\bigskip
\begin{center}
{\large\bf ACKNOWLEDGEMENTS} \\
\end{center}
LM was supported in part by NSF DMS-1712041.  GH was supported in part by NSF DMS-1712554. DW's research on swallows was supported most recently by NSF DEB 1242573. DF was supported in part by The Leon Levy Foundation. 
We thank the eBird participants for their contributions and the eBird team for their support. eBird was supported by the Wolf Creek Foundation, NASA (NNH12ZDA001N-ECOF), and the National Science Foundation (ABI sustaining: DBI-1356308; with computing support from CNS-1059284 and CCF-1522054). Thanks to Satish Iyengar for providing valuable feedback on an early draft of this paper.


\bibliographystyle{apalike}

\section*{Supplementary Material}
\section*{S.1 Cross Validation Study}

We now provide more details about the cross validation (CV) study referenced in Section 3 of the main text. We train a random forest with \texttt{mtry = 5} (not chosen by cross validation) and 500 trees, a $k$-nearest-neighbors (KNN) regression model with $k$ chosen from $\{5,7,..,21,23\}$, a 3-layer artificial neural network (ANN) with the number of neurons chosen from $\{15, 30, 100\}$ at each layer \citep{Bergmeir2012}, Linear Discriminant Analysis (LDA), Quadratic Discriminant Analysis (QDA), and a Generalized Additive Model (GAM) with degrees of freedom chosen from $\{1,6,11,...,21,26\}$ \citep{Hastie2017}. Finally, we train an elastic-net penalized logistic regression (GLMNet) model \citep{Friedman2010}, with weights $\alpha \in \{0,1\}$, (0 corresponds to the Lasso, $1$ corresponds to Ridge Regression), with cost parameter $\lambda \in \{0,.01, ..., .15\}$. This model fits coefficients to all covariates and also every two-way interaction, to parsimoniously select the strongest interaction models. These models are trained using the \texttt{caret} package \citep{Kuhn2017} in \texttt{R}, and, with the exception of random forests, the parameters chosen reflect those which lead to the smallest CV estimate of Root Mean Squared Error (RMSE). We also report the CV estimate of the Mean Absolute Error (MAE).

\begin{figure}[H]
\begin{floatrow}
\ffigbox{%
  \includegraphics[scale=0.575]{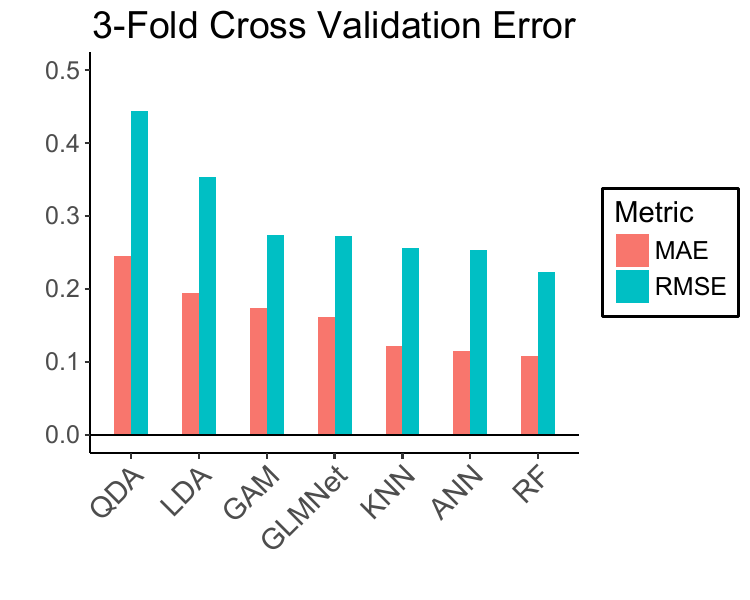}
}{%
  \caption{3-fold CV estimates of RMSE and MAE for various predictive models, plotted in descending RMSE order.}\label{fig:CV}
}
\capbtabbox{%
\begin{tabular}{|l|cc|}
  \hline
 Model& RMSE & MAE \\ 
  \hline
Random Forest \texttt{mtry = 5} & 0.22280 & 0.10863 \\ 
  ANN  (100, 30, 15) & 0.25300 & 0.11465 \\ 
  KNN ($k = 11$) & 0.25558 & 0.12195 \\ 
  GLMNet ($\lambda$ = 0) & 0.27260 & 0.16163 \\
  GAM  ($df =26$) & 0.27380 & 0.17412 \\  
  LDA & 0.35350 & 0.19400 \\ 
  QDA & 0.44427 & 0.24541 \\ 
   \hline
\end{tabular} 
}{%
  \caption{3-fold CV estimates of RMSE and MAE for various predictive models, tabulated in ascending RMSE order. Parameters listed correspond to those chosen by the CV analysis, except for the random forest \texttt{mtry} parameter.\label{tab:3foldCV} }
}
\end{floatrow}
\end{figure}
The results of this analysis are displayed in \autoref{fig:CV} and are tabulated in Table 1. Even without tuning, the off-the-shelf random forest model attains the lowest RMSE and MAE scores with the other flexible models, such as KNN and the ANN, not far behind. We see that the GAM and GLMNet models are similar in performance, with LDA and QDA lagging severely behind. Notably, the GLMNet model selected a tuning parameter that maximized model complexity, i.e. $\lambda = \alpha = 0$, even with all two-way interactions considered. This further suggests that a parametric model is unreasonable due to the complex interactions and functions of the covariates that go into predicting occurrence.  The strong predictive performance of random forests, combined with the recent advancements in inference for random forests, make them an ideal model for drawing conclusions about tree swallow migrations. As a first step analysis, we make use of traditional means for drawing inferences from black-box methodology of RFs with tools such as partial dependence plots and conclude with a spatial analysis of the predictive influence of \texttt{max\_temp}. These analyses provide heuristic and provisional answers to the study questions, and motivate the more formal testing procedures developed and executed in later sections.

\section*{S.3 A Causal Inference Analysis}

In keeping with the recommendations of one anonymous reviewer, we now implement an analysis to estimate the effect of \texttt{max\_temp\_anomaly} on \texttt{occurrence}. We note that the quantity of interest here is distinct from that in the main text, and so we begin with a brief overview of the potential outcomes framework and causal random forests. For a continuous treatment $W$, the average treatment effect at a point $x$ is defined as
\[
\tau(x) = \frac{\text{Cov}(Y, W | X = x)}{\text{Var}(W | X = x)}
\]
so that $\tau(x)$ measures the average \textit{linear} effect of $W$ on $Y$ given covariates $X = x$. This is an extension of the potential outcomes framework \citep{Rubin2005}, which defines the counterfactual treatments, $Y^{(w)}$ for each $w$ in the support of $W$. The fundamental goal of causal random forests is to estimate $\tau(x)$ \citep{Wager2017}. Causal trees, for continuous treatments,  proceed by recursively partitioning the feature space until some stopping criterion is met and then performing local linear regression of $Y$ on $W$ within the terminal nodes. Then, a prediction of $\tau(x)$ is given by the estimated treatment effect within the terminal node containing $x$. The forest is generated by repeatedly creating randomized trees, and averaging the estimated treatment effect from each tree. \citet{Wager2017} showed that if several regularity conditions are enforced upon training of the trees, then $\hat{\tau}(x)$ is asymptotically consistent to the true effect.

To interpret $\hat{\tau}(x)$ as an estimate of a causal effect, one must place an additional assumption on the distribution of the data, namely \textit{unconfoundedness}, which states that
\[
Y^{(w)}_i \ \indep \ W_i \ |\ X_i \quad \forall \quad w.
\]
Essentially, the response, given a particular treatment assignment $w$, needs to be locally (in $X$)  independent of the process by which treatments are assigned. The assumption can also be viewed as stating that the covariates $X$ are a sufficient \textit{adjustment set} to infer the causal effect of $W$ on $Y$. 

\begin{figure}[t]
\centering
\includegraphics[width = 0.45\textwidth]{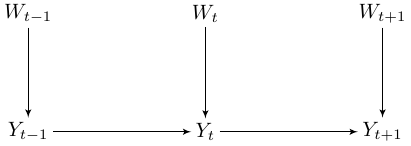} 
\vline \hspace{2mm}
\includegraphics[width = 0.45\textwidth]{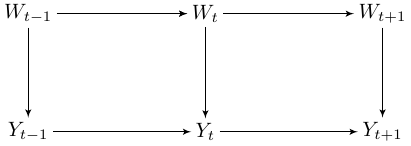}
\caption{Two different directed acyclic graphs (DAGs) describing the relationship between $W_t$ and $Y_t$. \textbf{Left:} A treatment scheme that would satisfy unconfoundedness. \textbf{Right:} A more likely DAG, for which unconfoundedness does not hold.} \label{fig:DAG}
\end{figure}

In our application, $Y$ is \texttt{occurrence}, $W$ is \texttt{max\_temp\_anomaly}, and $X$ are the remaining covariates such as land cover, time of year, user characteristics, etc. Thus, the unconfoundedness assumption translates to assuming that  the distribution of temperature anomalies is independent of the distribution of potential occurrences for all possible temperatures, conditional on the other covariates. We believe that the unconfoundedness assumption is likely unrealistic in this situation. Consider that both \texttt{occurrence} and \texttt{max\_temp\_anomaly} are both realizations of time series with high serial dependence. In \autoref{fig:DAG}, we present two different directed acyclic graph (DAG) representations of the time series structure of the data, fixed at a location $x$. In the left panel, if we assume that $W_{t}$ are sequentially independent, but perhaps day-to-day occurrences are dependent, then unconfoundedness holds. However, in the right panel, we model the more realistic scenario, where both the treatment and outcome are serially dependent, so that $Y_t^{(w_t)}$ and $W_t$ are confounded by $W_{t-1}$. As such, we have reason to doubt that $\tau(x)$ is identifiable from this data. Further, we do not observe $W_{t-1}$, and so cannot include it in an adjustment set. The problem becomes more intractable when one considers that the causal mechanism between $W$ and $Y$ is primarily insect activity, as described in the introduction of the main text, which acts as an unobserved variable. Because insect activity is \textit{also} serially dependent and unobserved, it can effectively make the adjustment set require infinite history of the time series observations, making causal inference impossible \citep{Malinsky2018}.

With these caveats in mind, we now apply the causal forest algorithm to the data used in Section 5 of the main text. We use the same training and test points described in the main text and apply the causal forest algorithm implemented in the \texttt{grf} package \citep{Athey2019} with the default parameters (including tree honesty) enabled. Then, predictions $\hat{\tau}(x)$ were recorded for each of the stratified test points. 

\begin{figure}[t]
\centering
\includegraphics[width = .45\textwidth]{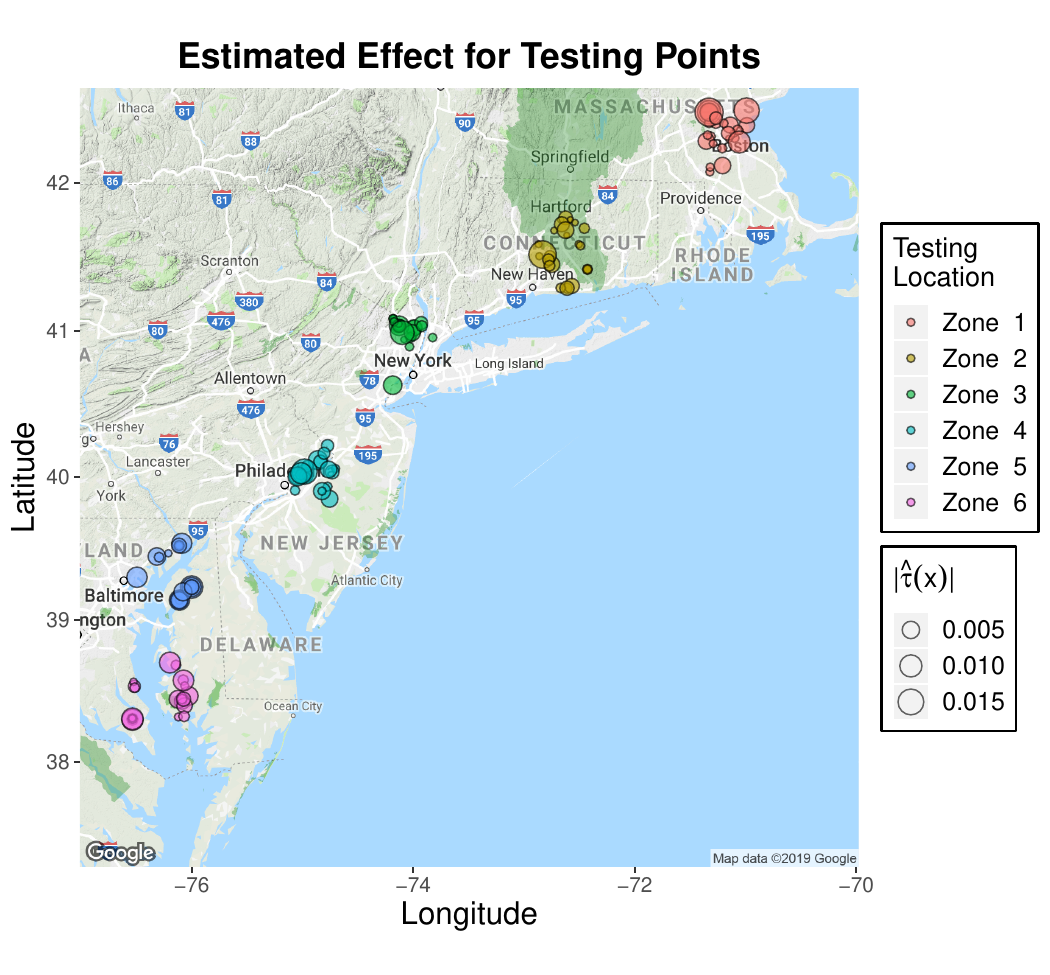}
\includegraphics[width = .45\textwidth]{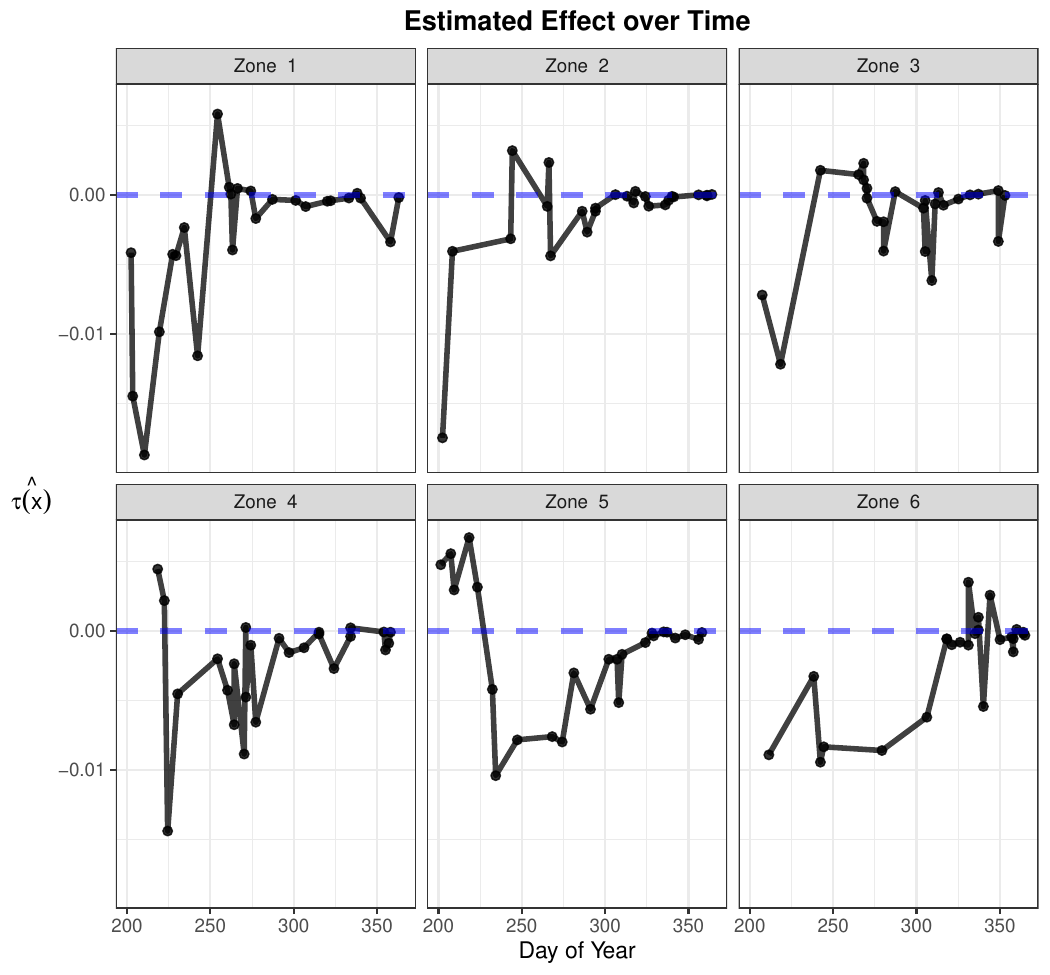}
\caption{\textbf{Left:} Absolute causal forest estimates $|\hat{\tau}(x)|$ for each point in the test set used in Section 5 of the main text. Size of the circle corresponds to magnitude of the estimated treatment effect. \textbf{Right:} A plot of $\hat{\tau}(x)$ over time in each zone.} \label{fig:MapTS}
\end{figure}

It is hard to discern any spatial trend in $|\hat{\tau}(x)|$ from the left panel of \autoref{fig:MapTS}. However, the time series plots tell an intuitive story - negative temperature anomalies lead to reduced occurrence earlier in the season (particularly in the the northern testing zones), followed by a flattening of the effect later in the migration season. Notice that the flattening occurs the earliest in Zone 1 (around \texttt{DoY} 270),  and latest in Zone 6 (around \texttt{DoY} 320), which again coincides with spatial differences in seasonality. We note that these results are in consensus with the formal testing from Section 4 of the main text, where differences between the regression functions died down later in the year.

\end{document}